\documentstyle[emulateapj]{article}
\singlespace
\begin{document}

\def\gsim { \lower .75ex \hbox{$\sim$} \llap{\raise .27ex \hbox{$>$}} }
\def\lsim { \lower .75ex \hbox{$\sim$} \llap{\raise .27ex \hbox{$<$}} }

\title{The cosmological origin of the Tully-Fisher relation}
\lefthead{Steinmetz \& Navarro}
\righthead{The cosmological origin of the Tully-Fisher relation}

\author{Matthias Steinmetz\altaffilmark{1}}

\affil{Steward Observatory, University of Arizona, Tucson, AZ 85721, USA}

\and 

\author{Julio F. Navarro\altaffilmark{2}}

\affil{Department of Physics and Astronomy, University of
Victoria, Victoria, BC V8P 1A1, Canada}

\altaffiltext{1}{Alfred P.~Sloan Fellow. Email: msteinmetz@as.arizona.edu}
\altaffiltext{2}{CIAR Scholar. Email: jfn@uvic.ca}

\begin{abstract}
We use high-resolution cosmological simulations that include the effects of
gasdynamics and star formation to investigate the origin of the Tully-Fisher
relation in the standard Cold Dark Matter cosmogony. Stars are assumed to form
in collapsing, Jeans-unstable gas clumps at a rate set by the local gas density
and the dynamical/cooling timescale. The energetic feedback from stellar
evolution is assumed to heat the gas surrounding regions of ongoing star
formation, from where it is radiated away very rapidly. The star formation
algorithm thus has little effect on the rate at which gas cools and collapses
and, as a result, most galaxies form their stars very early. Luminosities are
computed for each model galaxy using their full star formation histories and the
latest spectrophotometric models. We find that at $z=0$ the stellar mass of
model galaxies is proportional to the total baryonic mass within the virial
radius of their surrounding halos. Circular velocity then correlates tightly
with the total luminosity of the galaxy, reflecting the equivalence between mass
and circular velocity of systems identified in a cosmological context. The slope
of the relation steepens slightly from the red to the blue bandpasses, and is in
fairly good agreement with observations. Its scatter is small, decreasing from
$\sim 0.45$ mag in the U-band to $\sim 0.34$ mag in the K-band.  The particular
cosmological model we explore here seems unable to account for the zero-point of
the correlation. Model galaxies are too faint at $z=0$ (by about two magnitudes)
if the circular velocity at the edge of the luminous galaxy is used as an
estimator of the rotation speed.  The Tully-Fisher relation is brighter in the
past, by about $\sim 0.7$ magnitudes in the B-band at $z=1$, at odds with recent
observations of $z \sim 1$ galaxies. We conclude that the slope and tightness of
the Tully-Fisher relation can be naturally explained in hierarchical models but
that its normalization and evolution depend strongly on the star formation
algorithm chosen and on the cosmological parameters that determine the universal
baryon fraction and the time of assembly of galaxies of different mass.
\end{abstract}

\keywords{cosmology: theory -- galaxies: formation, evolution -- methods: numerical}

\section{Introduction}

The tight correlation between the total luminosity of disk galaxies and the
rotation speed of their gas and stars (Tully \& Fisher 1977) has been a
cornerstone of studies that attempt to measure galaxy distances and to map the
dynamics of the local universe (see, e.g., Strauss \& Willick 1995 for a
review). The substantial observational effort devoted to the subject over the
past two decades has provided a good understanding of the properties of this
relation in the local universe. The Tully-Fisher (TF) relation steepens
systematically from the blue to the red passbands and is surprisingly tight,
especially at longer wavelengths (the rms scatter is only $\sim 0.4$ mag in the
I-band, see Willick et al.~1997). The situation is less clear at higher redshift
where, depending on sample selection and the details of the observational
technique used to estimate rotation speeds, conflicting claims that the TF
relation either brightens or dims at modest redshifts ($z \lsim 1$) can be found
in the literature (Rix et al.~1997, Simard \& Pritchet 1996, Hudson et al.~1998,
Vogt et al.~1996, 1997).

From a theoretical point of view, ideas regarding the origin of the TF relation
can be divided into two broad camps: one that sees the TF relation as a result
of self-regulated star formation in disks of different mass (see, e.g., Silk 1997);
and another where the TF relation is a direct consequence of the cosmological
equivalence between mass and circular velocity (see, e.g., Mo, Mao \& White
1998). This equivalence is a consequence of the finite age of the universe, which
imposes a maximum radius from where matter can accrete to form a galaxy. This
radius (sometimes called the ``virial'' radius) corresponds roughly to that of a
sphere with mean density about $200$ times the critical density for closure
(White \& Frenk 1991, White et al.~1993). The circular velocity at the virial
radius ($V_{200}$) and its enclosed mass ($M_{200}$) then are equivalent measures of
the system mass related by
$$
M_{200}= 2.33 \times 10^5 \biggl({V_{200} \over {\rm km \,  s}^{-1}}\biggr)^3
\biggl({H_0 \over H(z)}\biggr) h^{-1} M_{\odot}. \eqno(1)$$
$H(z)$ is the value of Hubble's constant at redshift $z$, and $H_0=100 \, h$ km
s$^{-1}$ Mpc$^{-1}$ is its value at the present time.  Given this scaling, if
disk rotation speeds ($V_{\rm rot}$) are proportional to the circular velocity
of the halo ($V_{200}$) and stellar masses are proportional to $M_{200}$, then
luminosities should scale approximately like $V_{\rm rot}^3$, a velocity
dependence similar to that of the observed TF relation.

Both of these assumptions may fail in practice. Disk rotation speeds
depend in a non-trivial fashion on the stellar and gaseous mass of the galaxy,
on the contribution of the dark halo within the optical radius of the galaxy,
and on the radius where rotation speeds are measured. All these parameters may
vary significantly from galaxy to galaxy (see, e.g., Navarro 1998 and references
therein). Furthermore, luminosities, especially in the blue passbands, are not
simply proportional to the stellar mass, but rather reflect a subtle interplay
between the integrated and instantaneous star formation history of each
galaxy. Theoretical progress in this area then requires a careful evaluation of
these effects.

We present in this {\it Letter} a first attempt at addressing these issues
through direct numerical simulations of the formation of galaxies within their
full cosmological context. Our numerical experiments treat self-consistently
many of the relevant physical processes of the problem: gravity, gas dynamics,
and (in a simplified manner) the formation of stars and their energetic
feedback. The star formation history of each galaxy is used as input to the
latest spectrophotometric codes to compute broad-band luminosities that can be
readily compared with observations. Our simulations represent a significant
improvement over prior numerical work on the subject, which either neglected
star formation and feedback and focussed on the ability of gas the cool and
collapse within dark halos, or lacked the proper numerical resolution to analyze
the TF relation over a large range of mass and redshift (see, e.g., Evrard,
Summers \& Davis 1994, Navarro \& White 1994, Tissera, Lambas \& Abadi 1997).

\section{The Numerical Experiments}

The simulations were performed using GRAPESPH, a code that combines the Smoothed
Particle Hydrodynamics (SPH) approach to numerical hydrodynamics with a direct
summation N-body integrator optimized for the special-purpose hardware GRAPE
(Steinmetz 1996). GRAPESPH is fully Lagrangian and highly adaptive in space and
time due to the use of individual particle smoothing lengths and timesteps. It
is thus optimally suited to study the formation of highly non-linear systems
such as individual galaxy systems in a cosmological context.  The code used for
the simulations described in this paper include the self-gravity of gas, stars,
and dark matter, a full $3D$ hydrodynamical treatment of the gas, radiative and
Compton cooling, and a simple recipe for transforming gas into stars. 

The star formation algorithm we adopt is similar to that described in Steinmetz
\& M\"uller (1994, 1995, see also Katz 1992 and Navarro \& White 1993). Star
formation is modeled by creating new collisionless ``star'' particles in
collapsing regions that are locally Jeans-unstable at a rate given by
$\dot{\varrho}_{\star}=c_{\star} \varrho_{\rm gas}/\max(\tau_{\rm
cool},\tau_{\rm dyn})$. Here $\varrho_{\rm gas}$ is the gas density and
$\tau_{\rm cool}$ and $\tau_{\rm dyn}$ are the local cooling and dynamical
timescales, respectively. The proportionality parameter, $c_{\star}=0.05$, is
chosen so that most eligible gas is transformed into stars in about $\sim 20 \,
\tau_{\rm dyn}$ in dense regions, where $\tau_{\rm cool}\ll\tau_{\rm dyn}$. Our
results at $z=0$ are weakly dependent on our choice of $c_{\star}$ since
dynamical and cooling timescales in regions that actively form stars are much
shorter than the age of the universe.

After formation, ``star'' particles are only affected by gravitational forces,
but they devolve energy to their surroundings in a crude attempt to mimic the
energetic feedback from evolving stars and supernovae. Star particles inject
$10^{49}$ ergs per solar mass of stars formed into their surrounding gas $10^7$
yrs after their formation. This energy is invested into raising the internal
energy (temperature) of the gas. However, because stars form in high-density
regions, where cooling timescales are short, the feedback energy is almost
immediately radiated away, and as a result the star formation history traces the
rate at which gas cools and collapses to the center of dark matter halos. Our
implementation may thus be thought of as a ``minimal feedback'' model. We
caution that other possible implementations, in particular those that include
modifications to the kinetic energy of the gas surrounding star forming regions,
may lead to quite different star formation histories than the ones we report
here (Navarro \& White 1993).

The cosmological model we investigate is the standard Cold Dark Matter (CDM)
scenario, the paradigm of hierarchical clustering models of structure formation,
with the following choice for the cosmological parameters: $\Omega=1$, $h=0.5$,
$\Omega_{\rm b}=0.0125 \, h^{-2}$, and $\Lambda=0$. The power spectrum is
normalized so that at $z=0$  the rms amplitude of mass fluctuations in $8
h^{-1}$\,Mpc spheres is $\sigma_8=0.63$. Although long discredited as a viable
model for structure formation in the universe, the standard Cold Dark Matter
scenario is the best studied hierarchical clustering model and therefore it
serves as a useful cosmological testbed of different galaxy formation models.

We simulate spherical regions that evolve to form halos with circular velocities
in the range (80, 350) km s$^{-1}$. These regions are identified in cosmological
simulations of large periodic boxes and resimulated individually at higher
resolution, including the tidal field of the original simulation. Each region is
filled initially with $\sim 32,000$ gas particles and the same number of dark
matter particles. The size of the sphere is scaled in each case so that halos
have about the same number of particles at $z=0$. This ensures comparable
numerical resolution in systems of very different mass.  Details of the
procedure can be found in Navarro \& White (1994). The gas particle mass
varies between $2.5 \times 10^6 $ and $5 \times 10^7 h^{-1} M_{\odot}$,
depending on the mass of the system under consideration.  Dark matter particle
masses scale in the same way and vary from $5 \times 10^7$ to $ 10^9 h^{-1}
M_{\odot}$. These particle masses are small enough to exclude artificial heating
effects due to collisions between gas and dark matter particles (Steinmetz \&
White 1997). All simulations start at $z=21$ and use gravitational softenings
between $0.5$ and $1.25 h^{-1}$ kpc.

Stars form preferentially in collapsed, dense clumps of gas, and therefore model
``galaxies'' are easily identified in these simulations as dense collections of
star particles.  In the interest of simplicity, we retain for analysis only the
most massive galaxy inhabiting each individual halo. Most stars in a galaxy are
contained within a radius $r_{\rm gal}=15 \, (V_{200}/220$ km s$^{-1}) \,
h^{-1}$ kpc, which we use as a fiducial radius to compute all properties of the
luminous component of model galaxies. We also cull from the sample all ongoing
major mergers, defined as galaxies where the rms velocity of the stars differs
from the circular velocity at $r_{\rm gal}$ by more than $30 \%$.  In practice
less than $10 \%$ of the systems are eliminated this way from the analysis. The
final sample consists of $93$ galaxies at $z=0$ and of $44$ galaxies at $z=1$,
most of which contain more than $1,000$ star particles.

Galaxy luminosities are computed by simply adding up the luminosities of each
star particle, taking into account the time of creation of each particle (its
``age'') and its metallicity, as described in detail by Contardo, Steinmetz \&
Fritze-von Alvensleben (1998). The latest version of the spectrophotometric
models of Bruzual \& Charlot (1993) is also used for comparison.

\section{Results and Discussion}

\subsection{The I-band Tully-Fisher relation at $z=0$}

Figure 1 shows the simulated Tully-Fisher relation at $z=0$ in the I-band,
compared with the data of Giovanelli et al.~(1997), Mathewson, Ford \& Buchhorn
(1992) and Han \& Mould (1992). The lower panel shows I-band magnitudes versus
$V_{200}$, the circular velocity at the virial radius; the upper panel uses
instead $V_{\rm rot}$, the circular velocity measured at the fiducial radius
$r_{\rm gal}$.  The slope and scatter of the simulated TF relation are in fairly
good agreement with the observational data, regardless of which velocity
estimator is used. This result also holds in other bandpasses: the model TF
relation becomes shallower (and the scatter increases) towards the blue, just as
in observational samples (see Table 1).

The zero-point is also in good agreement with the data, provided that $V_{200}$
is used to estimate rotation speeds. Circular velocities at $r_{\rm gal}$ (the
edge of the luminous galaxies) are about $40$-$60 \%$ higher, so that adopting
$V_{\rm rot}$ as a velocity estimator results in a TF relation almost two
magnitudes too faint at given rotation speed. The large difference between
$V_{\rm rot}$ and $V_{200}$ is caused by the large fraction of baryons that
assemble into the central galaxy and the dark mass they draw in as they settle
at the center of the halo. On average, about $70$-$90\%$ of all baryons within
the virial radius are confined within $r_{\rm gal}$ at $z=0$.

\bigskip
\begin{tabular}{l|ccccc}
Band & $M_0 - 5\,\log(h)$ & $\delta_{M_0}$ & $A$ & $\delta_A$ & $\sigma(M)$\\
\hline
U & -18.50 & 0.066 & 6.95 & 0.30 & 0.45\\
B & -18.66 & 0.060 & 7.08 & 0.28 & 0.39\\
V & -19.34 & 0.058 & 7.20 & 0.25 & 0.35\\
R & -19.89 & 0.056 & 7.30 & 0.25 & 0.34\\
I & -20.27 & 0.060 & 7.33 & 0.25 & 0.34\\
K & -21.95 & 0.053 & 7.48 & 0.23 & 0.34\\
\end{tabular}

\medskip

\noindent{\small \bf Table 1:}
{\small Parameters of fits of the form $M = M_0 + A \log (V_{\rm
rot}/200\,\mbox{km\,s}^{-1})$ to the TF relation in different
bands. Errors in $M_0$ and $A$ are $(10,90)$ percentile bootstrap
deviations from the mean. $\sigma(M)$ is the rms scatter in magnitudes
of a least square fit.}\bigskip

This illustrates what is likely to be a fatal problem for this particular
cosmological model. Because dark halos formed in this scenario are quite
centrally concentrated (Navarro et al.~1997), the assembly of a massive galaxy
at the center raises $V_{\rm rot}$ above and beyond the halo circular velocity,
by up to $60 \%$ in the standard CDM scenario we investigate here. The effect of
this increase is a significant mismatch of the zero-point in the model TF
relation relative to observations. Model galaxies would have to be almost 2
magnitudes brighter in order to match the observed TF relation. This is highly
unlikely, since galaxies in our model contain most of the baryonic material
within the virial radius.

The only way to collect a massive disk galaxy without significantly increasing
$V_{\rm rot}$ over $V_{200}$ is to have dark halos which are less dense than
those formed in the standard CDM scenario. This favors low-density ($\Omega_0
<1$) universes, where dark halos are less centrally concentrated (Navarro et al
1997, Navarro 1998). In this case, the problem is further alleviated because the
universal baryon fraction is larger (by a factor $\Omega_0^{-1}$) and therefore
massive, bright galaxies can be assembled by collecting a relatively small
fraction of baryons into the central galaxy.

{\epsscale{0.4}
\plotone{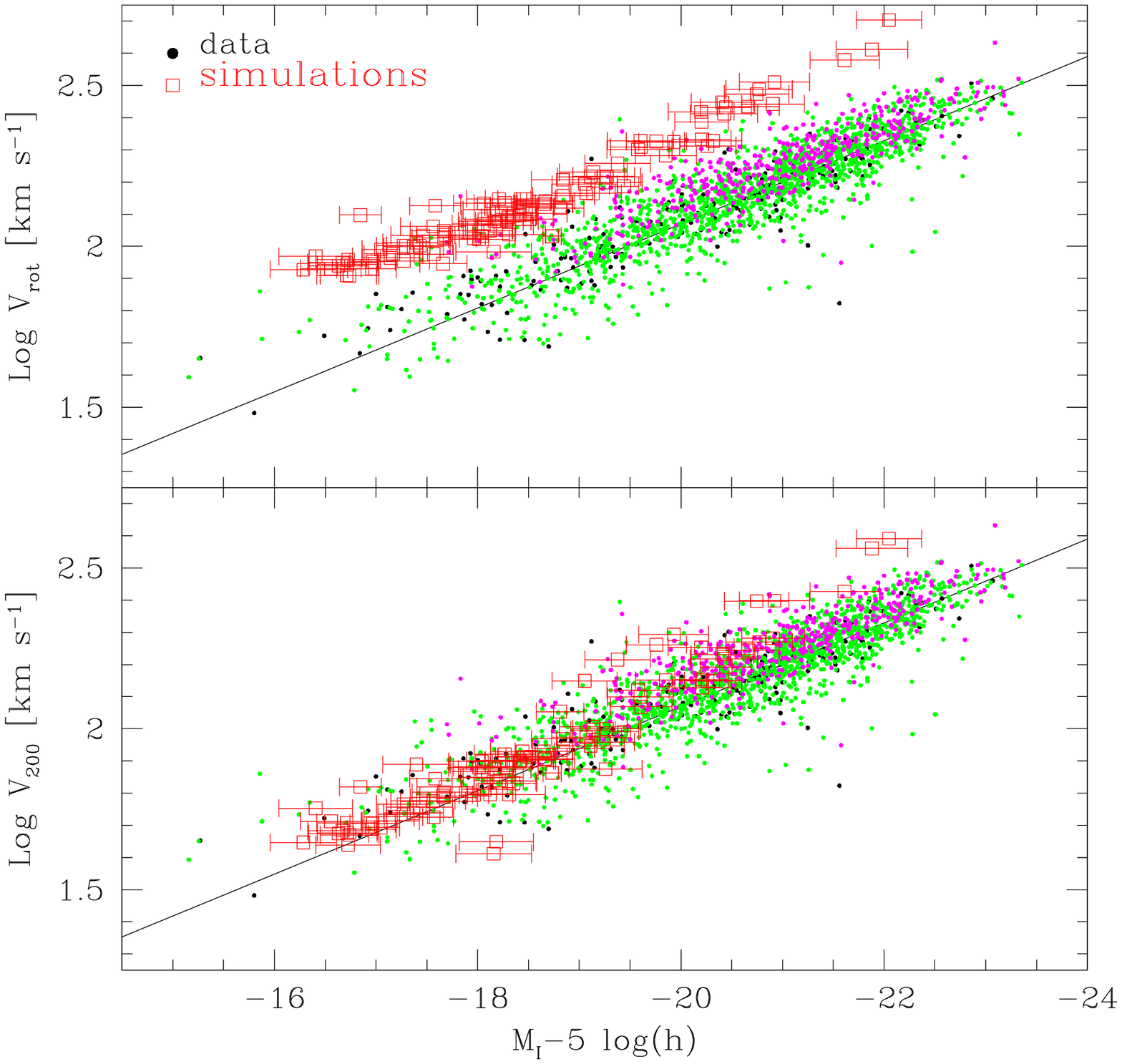}}

{\small {\sc Fig.}~1.---I-band Tully Fisher relation at $z=0$. The error bars in
the simulated data indicate the difference in magnitudes that results from
adopting a Salpeter or a Scalo IMF.  }\bigskip

\subsection{The star formation history of model galaxies}

As discussed in \S2, our feedback implementation has minimal impact on the rate
at which gas cools and collapses to the center of dark matter halos. The star
forming history of a model galaxy thus traces the history of cooling and
accretion of gas within halos, punctuated by small ``bursts'' of star formation
that coincide with major mergers. This is shown in Figure 2, where we plot the
combined star formation rate of four model galaxies, spanning a factor of $\sim
3$ in circular velocity. The large peak at early times is due to the nearly
simultaneous collapse of a number of halos which later merge to form the final
galaxy. The star formation rate in each progenitor rarely exceeds $\sim 50
M_{\odot}$/yr, although its combined rate can be much higher.

Most simulated galaxies form their stars quite early (half of them typically by
$z\sim 1$-$1.5$), and have almost exhausted their gas supply at $z=0$, when the
star formation rate is only about $20$-$30 \%$ of the global rate averaged over
the age of the universe. These star formation histories are thus similar to
those deduced for E/S0/Sa galaxies rather than to the nearly constant star
formation rates characteristic of late-type spirals (Kennicutt 1998).

\bigskip
{\epsscale{0.45}
\plotone{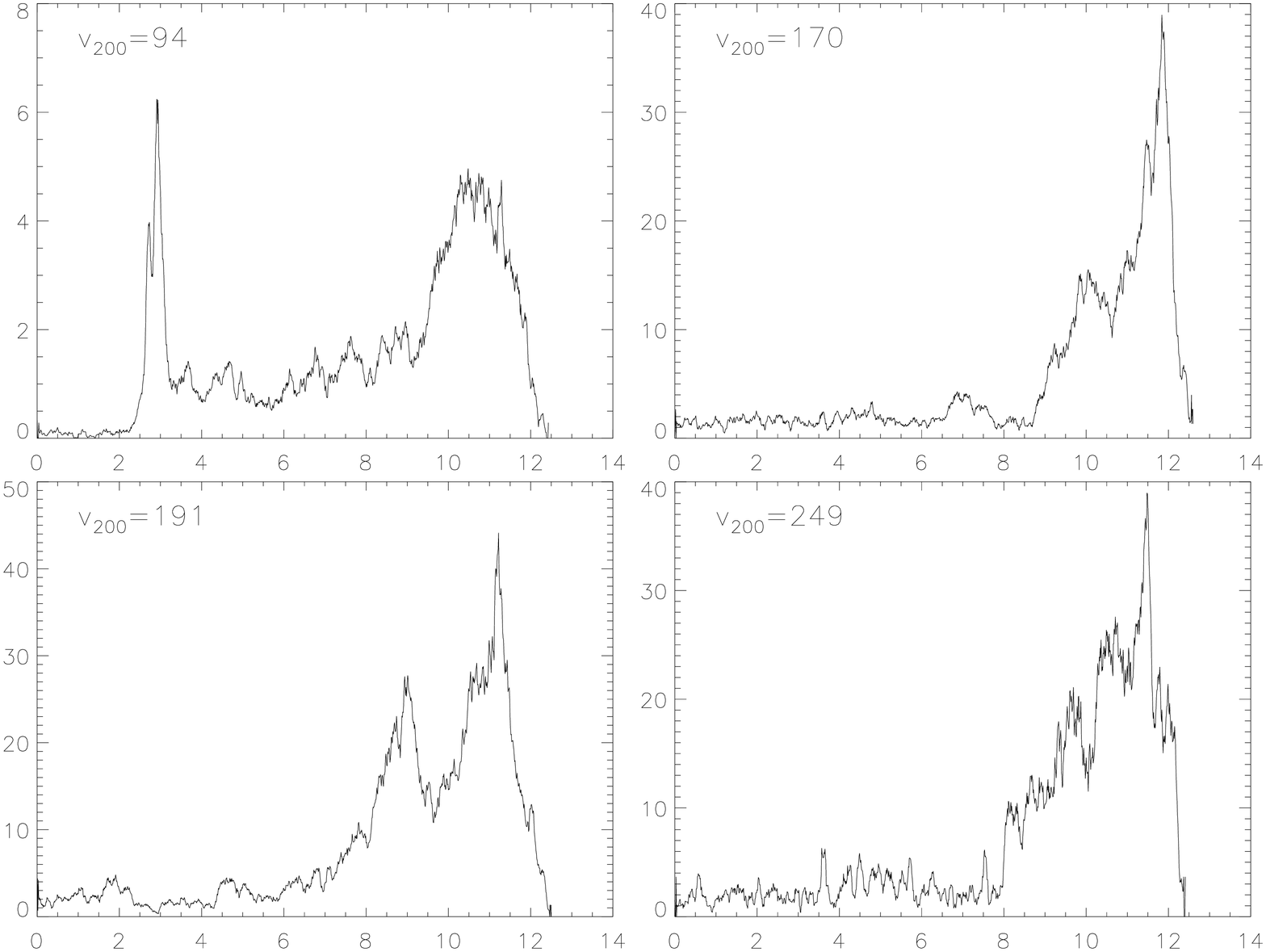}}

{\small {\sc Fig.}~2.---Star formation rate smoothed over $10^8$ yr intervals
(in solar masses per year) versus lookback time (in Gyr) for 4 typical halos
with virial velocities between 90 and 250\,km s$^{-1}$.}\bigskip

\subsection{The angular momentum of simulated galaxies}

As discussed in the previous subsection, most gas cools and collapses before the
merger events through which the galaxy is finally assembled. We therefore expect
the angular momentum of the model galaxies to be significantly lower than those
of disk galaxies, since angular momentum is transferred from the gas to the dark
matter during mergers (Navarro, Frenk \& White 1995, Navarro \& Steinmetz 1997).

This is shown in Figure 3, where we compare the specific angular momentum of the
model galaxies to the disk galaxies in the samples of Courteau (1997), Mathewson
et al.~(1992), and the compilation of Navarro (1998). Disk angular momenta are
estimated assuming that they are pure exponential systems with flat rotation
curves, $j_{\rm disk} = 2 \, V_{\rm rot} \, r_{\rm disk}$, where $V_{\rm rot}$
is identified with the (maximum) rotation speed and $r_{\rm disk}$ is the
exponential scalelength. The solid line shows, for comparison, the typical
specific angular momentum of dark halos, $j_{\rm halo}=(2/f_c)^{1/2} \, \lambda
\, (V_{200}/$km s$^{-1})^2$ km s$^{-1} h^{-1}$ kpc, where $f_c$ is a factor of
order unity that depends on the internal structure of halos (Mo et al.~1997) and
$\lambda$ is the usual dimensionless spin parameter, typically found to be
$\sim 0.05$ in cosmological N-body simulations (see, e.g., Cole \& Lacey
1996). (We assume $V_{200}$=$V_{\rm rot}$ to plot the halo curve.) Disk angular
momenta scale with rotation speed just as $j_{\rm halo}$ scales with circular
velocity, suggesting that baryons in disks have retained a roughly constant
fraction of the angular momentum of their surrounding halos: about one half if
$V_{200}=V_{\rm rot}$ (see Syer, Mo \& Mao 1998 for a more detailed discussion).

The simulated galaxies, on the other hand, have specific angular momenta which
are more than one order of magnitude lower than those of observed disks. This
confirms that the hierarchical assembly of a galaxy robs the baryonic component
of much of its angular momentum if gas cooling and star formation precedes most
major mergers. This angular momentum problem signal that feedback should be more
effective at preventing gas from collapsing and cooling in the early stages of
the hierarchy in order to ensure that most of the angular momentum is not lost
as the luminous component of a galaxy is assembled (see also Weil, Eke \&
Efstathiou 1998).

{\epsscale{0.4}
\plotone{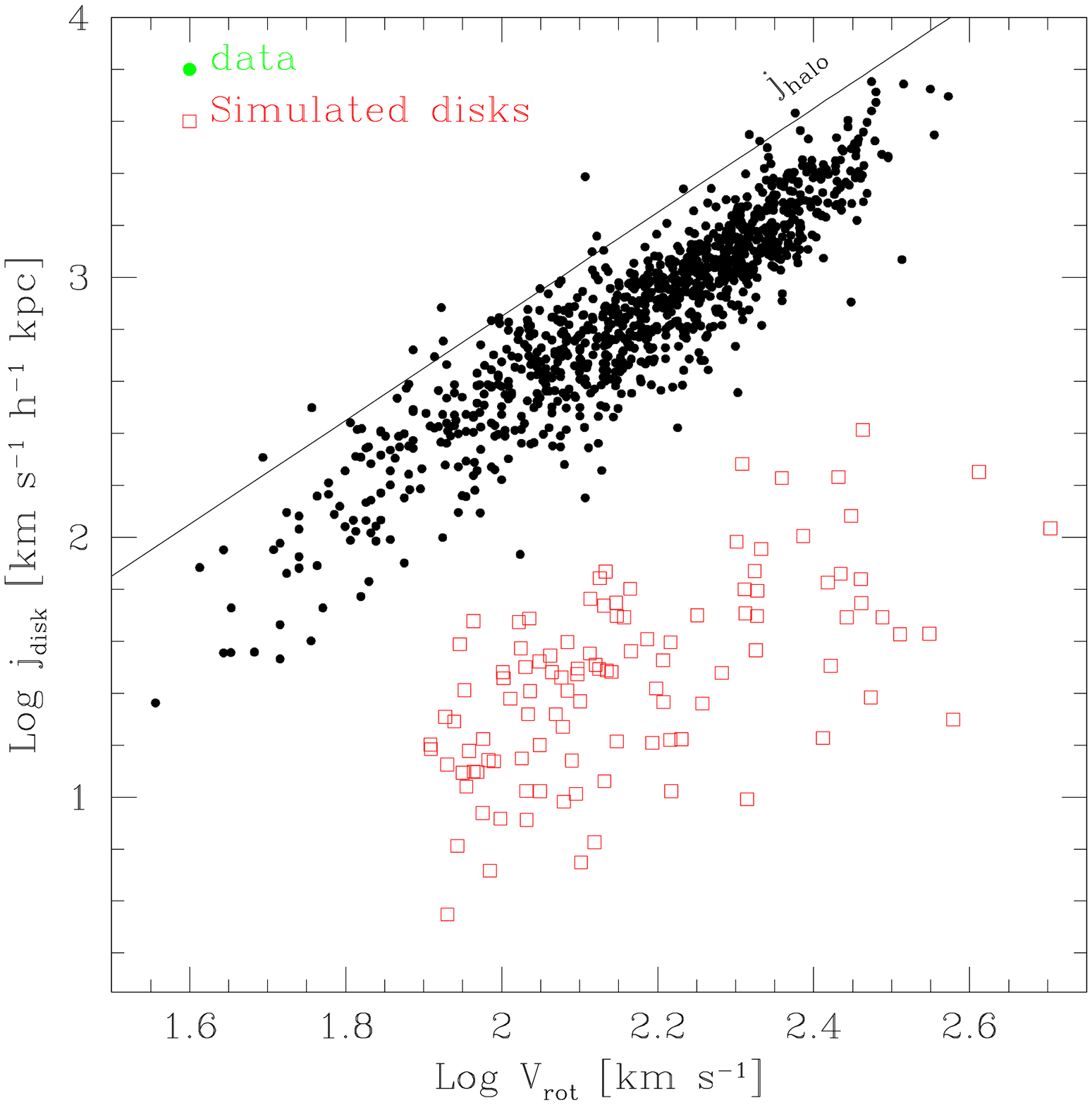}}

{\small {\sc Fig.}~3.---Specific angular momentum vs circular velocity of
model galaxies compared to observational data.}\bigskip
%

\subsection{The redshift evolution of the Tully-Fisher relation}

Figure 4 shows the redshift evolution of the TF relation in the rest-frame
B-band. The shaded region corresponds to the results at $z=0$, as reported by
Pierce \& Tully (1992). Starred symbols show the $z\sim 1$ data of Vogt et 
al.~(1996, 1997). All data have been scaled to the value of $q_0=0.5$ used in this
paper. Note that for this value of $q_0$ the data from Vogt et al.~are consistent
with a slight dimming of the TF relation at $z\sim 1$. Their claim of a modest
brightening holds only for the low value of $q_0=0.05$ adopted in their
analysis.

At $z=0$ the results for the B-band TF relation of simulated galaxies are
similar to those presented in Figure 1 for the I-band. The slope and scatter are
in good agreement with observations, but the zero-point depends strongly on the
radius at which circular velocities are measured. Good agreement is found
provided that $V_{200}$ is used as a measure of the rotation speed, but a
significant zero-point offset is observed if $V_{\rm rot}$ is used instead.

Independent of which velocity estimator is adopted, the model TF relation
brightens at $z=1$, by $\sim 0.7$ or $\sim 0.2$ mag, depending on whether
$V_{\rm rot}$ or $V_{200}$ is used, respectively. This is at odds with the
observational results and perhaps a bit surprising since masses scale like
$H(z)^{-1} \propto (1+z)^{-3/2}$ at fixed circular velocity (eq.~1) and are,
therefore, a factor of $\sim 2.8$ times smaller at $z=1$. However, star
formation rates are much higher at $z=1$ (by a factor of $\sim 4$) and the
decrease in mass is more than compensated by the higher abundance of young stars
present at that time.

This interpretation is confirmed by results in other passbands less sensitive to
the instantaneous star formation rate: the K-band TF relation actually dims by
about $0.1$-$0.4$ mag between $z=0$ and $1$ (depending again on whether
$V_{200}$ or $V_{\rm rot}$ is used). The brightening trend observed in Figure 4 is
therefore a direct consequence of the star formation algorithm we use, and not a
genuine feature of hierarchical clustering models. One example is provided by
the study of Mao, Mo \& White (1998), who use different assumptions for the
star formation history of disks, and argue for a roughly constant TF zero-point
between $z=0$ and $1$ in a cosmological model similar to the one we adopt here.

\bigskip
{\epsscale{0.4}
\plotone{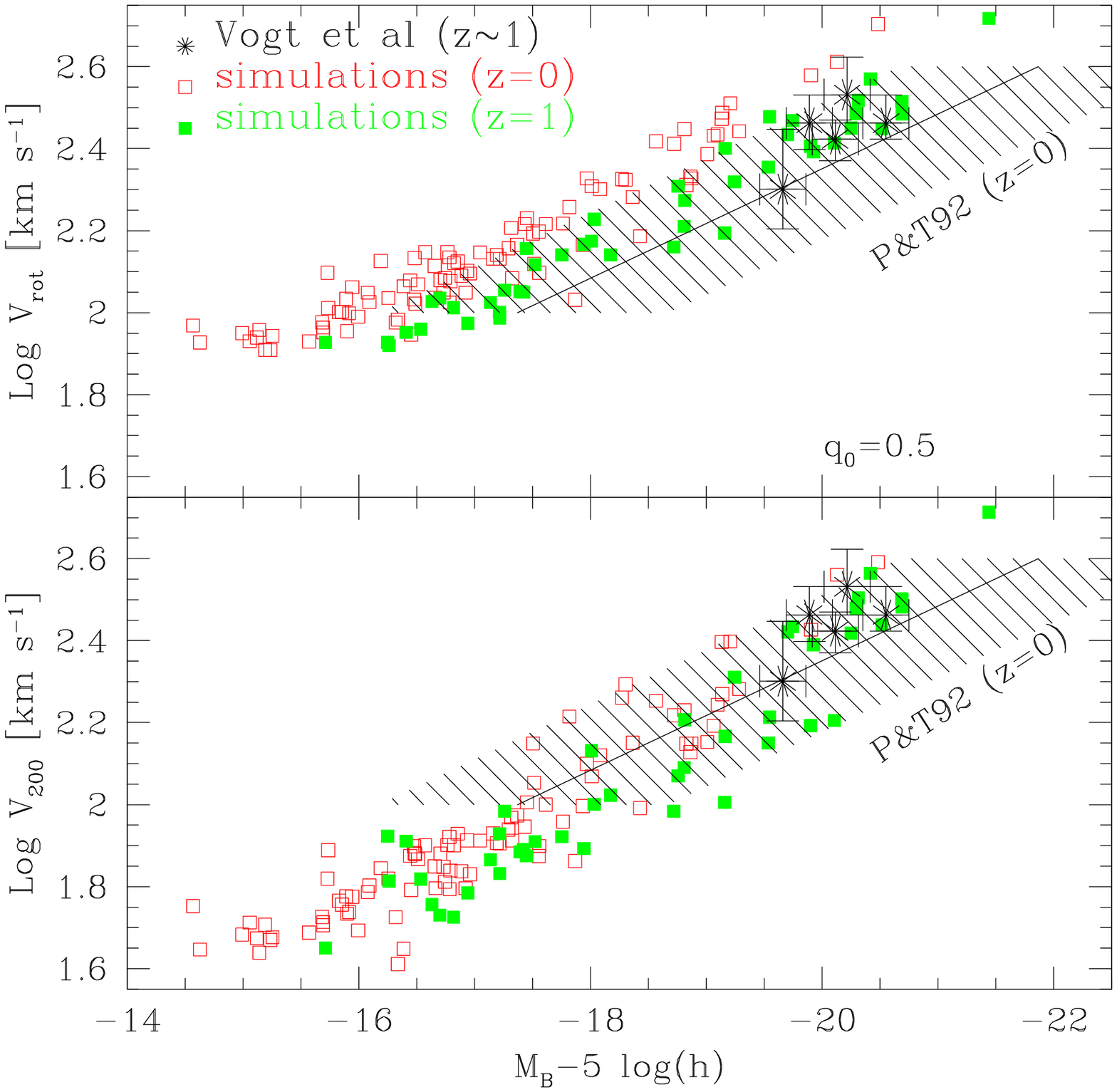}}

{\small {\sc Fig.}~4.---Redshift evolution of the rest-frame B-band
Tully-Fisher relation.}\bigskip

\section{Summary}

We have used high-resolution cosmological simulations which incorporate the
effects of gas pressure, hydrodynamical shocks, radiative cooling, star
formation and supernova feedback to study the origin of the TF relation of
spiral galaxies. Our main conclusion is that the slope (and scatter) of this
relation is naturally reproduced in hierarchically clustering universes, and
reflect the equivalence between mass and circular velocity of systems formed in
a cosmological context. The particular model we explore here (the standard CDM
scenario) seems unable to account for the zero-point of the correlation. Model
galaxies are too faint at $z=0$ (by about two magnitudes) if the circular
velocity at the edge of the luminous galaxy is used as an estimator of the
rotation speed. This result had already been hinted at, although using different
arguments, by semianalytic models of galaxy formation (Kauffmann, White \&
Guiderdoni 1993, Cole et al.~1994, but see Somerville \& Primack 1997 for a
dissenting view). 

The model TF relation brightens in the past, at odds with current observational
evidence. However, this is a direct result of the particular star formation
algorithm we use, which transforms gas into stars very efficiently at high
redshift.  Efficient gas cooling (and star formation) that precedes the final
assembly of the galaxy through mergers leads to the formation of stellar systems
with angular momenta far lower than those of observed disks. The combination of
successes and failures we report here demonstrate the potential of galaxy
scaling laws and their evolution as probes of the intricate physics of star
formation and feedback operating in a cosmological context.

\acknowledgments This work has been supported by the National Aeronautics and
Space Administration under NASA grant NAG 5-7151 and by NSERC Research Grant
203263-98. MS was supported in part by a fellowship of the Alfred P.~Sloan
Foundation. We thank Stephane Courteau, Nicole Vogt, Riccardo Giovanelli and the
MarkIII collaboration for making their data available in electronic form, and
Cedric Lacey for making available his compilation of the Mathewson et al.~data.

\end{document}